\documentclass[showpacs,preprintnumbers,amsmath,amssymb]{revtex4}
\usepackage{graphicx}
\begin{document}

\preprint{}

\title{Counting the negative eigenvalues of the thermalon in three dimensions}

\author{Rodrigo Aros} \affiliation{Departamento de Ciencias fisicas,\\ Universidad  Andr\'es Bello, Av. Rep\'ublica 239, Santiago, Chile.}

\author{Andr\'es Gomberoff} \affiliation{Departamento de Ciencias fisicas,\\ Universidad  Andr\'es Bello, Av. Rep\'ublica 239, Santiago, Chile.}

\author{Alejandra Montecinos} \affiliation{Departamento de F\'isica, Facultad de Ciencias F\'isicas y Matem\'aticas, Universidad de Chile, Casilla 487-3, Santiago, Chile.}

\begin{abstract}
Some years ago it was shown that the cosmological constant may be reduced by thermal production of membranes that, after nucleation, collapse into a black hole. The probability of the process was calculated in the leading semiclassical approximation by studying an associated Euclidean configuration called the thermalon. Here we investigate the thermalon in three spacetime dimensions, describing the nucleation of closed strings that collapse into point particle singularities. In this context we may analyze the one-loop structure without the well known problems  brought in by the propagating gravitational degrees of freedom. We found that the coupling to gravity may increase the number of negative eigenvalues of the operator.

\end{abstract}


\maketitle

%%%%%%%%%%%%%%%%%%%%%%%%%%%%%%%%%%%%%%%%%%%%%%%%%%%%%%%
%%%%%%%%%%%%%%%%%%%%%%%%%%%%%%%%%%%%%%%%%%%%%%%%%%%%%%%
\section{Introduction}
%%%%%%%%%%%%%%%%%%%%%%%%%%%%%%%%%%%%%%%%%%%%%%%%%%%%%%%
%%%%%%%%%%%%%%%%%%%%%%%%%%%%%%%%%%%%%%%%%%%%%%%%%%%%%%%

Currently some of the major challenges of theoretical physics are related to the cosmological constant. First, the infamous \textit{cosmological constant problem}, in which the observed cosmological constant is smaller in at least 60 orders of magnitude than the estimated value \cite{cosmoproblem}. Second, there is the question of why the cosmological constant was big at the early stages  of cosmic history, producing inflation, and how it relaxed to a very tiny number transferring its energy into normal matter. Third, the universe seems to be accelerating \cite{obs}, which is normally explained by the presence of a positive cosmological constant or, more generally, some kind of (unknown) dark energy whose nature has escaped our present knowledge.

In this work we deal with the first two problems. Specifically, we continue the study started in \cite{GHTW} on the decay of the cosmological constant by thermal nucleation of bubbles of true vacuum. This process may be described by the so called \textit{thermalon}, an Euclidean, time independent, solution of a charged membrane coupled to gravity and, in $D$ dimensions, a $(D-1)$-form electromagnetic potential. The nucleation rate of this process is of the form \begin{equation} \Gamma = A e^{-B}, \label{rate} \end{equation} where  $B=I_{back}-I_{therm}$, $I_{therm}$ is the action of the system evaluated on the thermalon, and $I_{back}$ is the action evaluated in the false vacuum background. Here we use the same convention used in  \cite{GHTW}, in which the Euclidean path integral includes $e^{I}$. The prefactor $A$ corresponds to the one loop contribution to the decay rate, namely, it is determined by the determinant of the operator ${\cal O}$ that arises from the second order perturbations of the action \begin{equation}\label{oper} \delta^{(2)} I_{therm} =  -\frac{1}{2} \int d^3 x \psi {\cal O} \psi.
\end{equation}
Here $\psi$ stands for the independent degrees of freedom of the action.

Unfortunately, to compute the above expression is in general extremely difficult. First, one must face the inherent difficulties in computing functional determinants. Second, gravity is not renormalizable for $d>3$. This second problem may be avoided by analyzing the thermalon solution in three spacetime dimensions, which is what we do in what follows. We expect that this toy model may shed some light on the problem of the negative eigenvalues around the thermalon. Depending on the classical solution considered the operator $\mathcal{O}$ may have any number of negative eigenvalues. On the other hand, on some general grounds, Coleman proved \cite{Coleman} that a classical solution which would yield an operator $\mathcal{O}$ with more than a single negative value ``has nothing to do with decay rates". Nonetheless, that demonstration manifestly excludes de Sitter spaces, and therefore there is no objection in principle to extend the analysis to operators with a larger number of negative eigenvalues. There is some evidence, however, that Coleman's result may be extended to a more general family of theories, including, in particular, field theory in dS backgrounds (see \cite{gravbounces}). We thus expect that if the thermalon is going to describe a decay process, it should have one and only one negative eigenvalue.
Unfortunately, it is known that in some cases with fixed background  the number of negative eigenvalues should be at least five\cite{urbano}. The immediate question is whether the introduction of the gravitational degrees of freedom may reduce the number of negative eigenvalues. We explore this possibility in this work and conclude that, under some general assumptions, gravity actually increases the number of negative eigenvalues.

The plan of the article is as follows. In section \ref{review} we review the thermalon general scheme and describe the three-dimensional case. Next, in section \ref{fixedback} we analyze the eigenvalues of the operator describing second order perturbations neglecting the back reaction of the geometry. Then, in section \ref{main} we analyze  the decay rate to first loop including the gravitational degrees of freedom. Finally we draw out some conclusions.

%%%%%%%%%%%%%%%%%%%%%%%%%%%%%%%%%%%%%%%%%%%%%%%%%%%%%%%%%%%%
%%%%%%%%%%%%%%%%%%%%%%%%%%%%%%%%%%%%%%%%%%%%%%%%%%%%%%%%%%%%
\section{The Thermalon in Three Dimensions} %%%%%%%%%%%%%%%%%%%%%%%%%%%%%%%%%%%%%%%%%%%%%%%%%%%%%%%%%%%
%%%%%%%%%%%%%%%%%%%%%%%%%%%%%%%%%%%%%%%%%%%%%%%%%%%%%%%%%%%%
\label{review}

We are interested in the following action, \begin{equation}\label{accion1} I_{E}=\frac{1}{2\kappa}\int_{\mathcal{M}} d^3x\sqrt{g} (R-2\lambda) + \frac{1}{12}\int_{\mathcal{M}} d^3x\sqrt{g}F_{\mu\nu\rho} F^{\mu\nu\rho}  - \int_{\partial \mathcal{M}} {}^*\! F\wedge A - q \int_{\Sigma} A -\mu\int_{\Sigma} \sqrt{h}d^2x, \end{equation} where $\mathcal{M}$ is the bulk and $\Sigma$ the worldsheet swept by a closed string of charge $q$ and tension $\mu$. The electromagnetic potential $A$ is a $2$-form and $F=dA$. The third integral is along the boundary of the manifold, and it is necessary to have as boundary condition the fixing of $F$. Finally, we have defined $\kappa=8\pi G$ and units in which $4\pi G=1$ such that $\kappa=2$.

The thermalon discussed in this work is a static, spherically symmetric extreme of action above (\ref{accion1}), therefore we will consider metrics of the form \begin{equation} \label{metricatermalon} ds^2_\pm=h^2_\pm(r)dt^2+f^{-2}_\pm(r)dr^2+r^2d\phi^2,
\end{equation}
where $0<t<\beta$, $0<\phi<2\pi$.  The subindices $\pm$ indicate the two regions of spacetime separated by the worldsheet of the string. The worldsheet may be parameterized such that $r=R(t,\phi)$. In this case, however, we are searching for a static and axisymmetric solution, thus $R(t,\phi)=R_{0}$ (constant). Since $F$ is a three form in three dimensions necessarily $F_{\mu\nu\rho}=\sqrt{g}E\epsilon_{\mu\nu\rho}$, where $E$ is a scalar, allowing to operate in term of the scalar $E$. For instance the electromagnetic equations yield \begin{equation}\label{maxwell} E_{+}-E_{-}=q .
\end{equation}
On the other hand, the Einstein equations have a matter term coming from the electric field which just adds a piece to the cosmological constant. At each region we have \begin{equation}\label{cc} \Lambda_\pm= \lambda + 4\pi G E^2_\pm \equiv \frac{1}{l_\pm^2}.
\end{equation}
To simplify the calculations, we are going to assume $\lambda=0$, such that the cosmological constant come entirely from the electromagnetic field.

Considering the restrictions above the Einstein equations imply that at each side of the string the geometry is locally de Sitter. The (+)-region  will be assumed to be free de Sitter space. The (-)-region will be the geometry of a particle singularity \cite{DJ} placed at $r=0$, \begin{equation} \label{Ds} ds^{2}_\pm=C^{2}_\pm f^{2}_\pm (r)dt^{2}+f^{-2}_\pm (r)dr^{2}+r^{2}d\phi^{2}, \end{equation} with \begin{equation} f^{2}_{+}(r)=1-\frac{r^2}{l^{2}_{+}}\textrm{, }f^{2}_{-}(r)=1- \frac{M}{\pi} -\frac{r^{2}}{l^{2}_{-}}.
\end{equation}
Here $M$ is the mass of the particle and $C_\pm$ are arbitrary constants, whose value is determined by the Israel matching conditions \cite{Israel}, which in this case require $g_{00}$ to be continuous across the string, hence $C_{-}=f_{+}(R)/f_{-}(R)C_+$.

Finally, one can always set the time coordinate, $t$, such that $C_+=1$ and therefore the killing vector $\partial_t$ be unitary at $r=0$.

The remaining Israel matching conditions  give \begin{equation} \label{I1} f_+(R)-f_-(R)=2\mu R, \end{equation} while the fact that we are searching for a static solution ($\dot{R}=\ddot{R}=0$) yields \begin{equation}\label{I2} f'_{+}(R)-f'_{-}(R)=2\mu.
\end{equation}

The two unknown quantities above, $R_{0}$ and $M$, can be determined by solving equations
(\ref{I1}) and (\ref{I2}). The radius of the thermalon is given by \begin{equation} \label{radio} R_{0}=\sqrt\frac{l_+^2}{1+x^2}, \end{equation} where $x =\gamma+\sqrt{1+\gamma^2}$, $\gamma=l_+\left((2\mu)^2-\alpha^2\right)/4\mu$ and $\alpha^2=1/l_+^2-1/l_-^2$. Analogously, the mass, $M$, is given by \begin{equation}\label{mass} M = \frac{2\pi\mu l_+}{x}.
\end{equation}

The geometry in the $(-)$-region is that of the particle singularity  introduced in \cite{DJT,DJ}.
To observe that we can perform the change of variables $r= (\sqrt{1-M/\pi}) \tilde r $, $\varphi= (\sqrt{1-M/\pi})\phi $ and observe that close to $r=0$ the geometry of a slice at constant time is given by $$ d\tilde{r}^2 + \tilde{r}^2 d\varphi , $$ with $0<\varphi<2\pi(\sqrt{1-M/\pi})$. This corresponds to a conical singularity of deficit $\sqrt{1-M/\pi}$. From (\ref{mass}) one obtains that $0<M/\pi<1$, which is the physically admissible interval\cite{DJ} ($M/\pi>1$ corresponds to sources of negative energy-density). In fact, from the definition of $\gamma$ one gets $$ 1 + \gamma^2 - \frac{l_+^2}{l_-^2} = (2\mu l_+
-\gamma)^2 , $$ and therefore $$ (1+\gamma^2) > (2\mu l_+ -\gamma)^2 .
$$
It is also clear form the definition that $2\mu l_+ >\gamma$, and therefore we obtain what we need, that is, $$ \sqrt{1+\gamma^2} + \gamma = x > 2\mu l_+ .
$$

The value of $B$ in Eq.(\ref{rate}) can be readily computed using the Hamiltonian formulation.
This is done in the appendix \ref{ap1} where we find that, on the constraint surface, the action is given by,

\begin{equation}\ \label{AccionOnMaxwell} I=-\mu \int_{\Sigma} \sqrt h d^2\sigma-\int_{\mathcal{M}} \sqrt g E^2 d^3x +\int_{\partial \mathcal{M}}{}^*\! F\wedge A +\pi {\cal A}+\frac{1}{2}\int_{\partial \mathcal{M}}(N\gamma_{,\rho} - 1).
\end{equation}

Now,
\begin{equation}\label{III} \int_{\partial \mathcal{M}}{}^*\! F\wedge A = E_+\int_{\partial \mathcal{M}} A = E_+ \int_{\mathcal{M}} F = E_+^2\int_{\mathcal{M}} \sqrt{g} = E_+^2 2\pi\beta_-\int_0^{l_+} r dr = \pi \beta_-, \end{equation}

where \begin{equation} \beta_-=2\pi l_-\sqrt{1-M/\pi} \end{equation} is the inverse temperature associated to the cosmological horizon of the $(-)$-region.  We  obtain a final expression for $B$ given by
\begin{equation}
B=-I+I_{Back}=\pi(\beta_+ - \beta_-)
\end{equation}
where $\beta_+=2\pi l_+$ is the temperature of horizon of the $(+)$-region.

Recalling  that $e^{-B}$ stands for the tree level semiclassical approximation to the probability of nucleating a string through thermal fluctuations we have found out the final expression to be evaluated.

Assuming that $q>0$ we notice, from Eq.(\ref{maxwell},\ref{cc}), that $l_{-}>l_{+}$ thus the cosmological constant is reduced in the region $r>R$ when the string materializes. After the  nucleation the system may be analyzed classically. In the Lorentzian sector, the thermalon is unstable, and may collapse both towards $r=r_{--}$ or towards  $r=0$.  In the first scenario, the (+)-region grows to cover the entire space  so that the initial situation is recovered. In the second one, the string collapses into a point particle \cite{DJ} and the (-)-region fills the entire spacetime. This is analogous to what happens in four-dimensional spacetime where the final geometry  is a Schwarzschild-de Sitter geometry with a smaller cosmological constant. More details on that process in four dimensions may be found in \cite{GHTW}.

%%%%%%%%%%%%%%%%%%%%%%%%%%%%%%%%%%%%%%%%%%%%%%%%%%%%%%%%%%%%%%%%%%%%%
%%%%%%%%%%%%%%%%%%%%%%%%%%%%%%%%%%%%%%%%%%%%%%%%%%%%%%%%%%%%%%%%%%%%%
\section{Thermalon in fixed curved background. Analysis of prefactor.} \label{fixedback} %%%%%%%%%%%%%%%%%%%%%%%%%%%%%%%%%%%%%%%%%%%%%%%%%%%%%%%%%%%%%%%%%%%%%
%%%%%%%%%%%%%%%%%%%%%%%%%%%%%%%%%%%%%%%%%%%%%%%%%%%%%%%%%%%%%%%%%%%%%

We will start by analyzing the operator ${\cal O}$ in (\ref{oper}) for the case in where the string back reaction in the gravitational field is neglected, \textit{i.e.}, in the limit $G\rightarrow 0$. The first consequence of this is that separation into $\pm$ sectors becomes irrelevant since neither the geometry nor the cosmological constant jump across the string, thus $l_{+}=l_{-}=l$.

Unlike the geometry the electromagnetic field jump across the -charged- string in concordance with the Gauss constraint (\ref{maxwell}). However in this case the electromagnetic field does not incorporate degrees of freedom since it is fixed between $r=0$ and the position of the string and determined by (\ref{maxwell}) \textit{behind} the string. Therefore the only degrees of freedom are those of the string.

Subtracting the background to the action (\ref{accion1}) we find \begin{equation}\label{accion} I = I_{back} - I_{therm} = \frac{\epsilon}{2}\int^{l}_{R} \sqrt g d^3x - \int_{\Sigma} \mu\sqrt h d^2\sigma, \end{equation} where $\epsilon=E_+^2 - E_-^2$. The first integral is between the radius of the string $R(t,\phi)$ and the cosmological horizon $l$. In the second term $h_{ij}$ is the induced metric on the string. For simplicity we choose the coordinates $\sigma=(t,\phi)$ on the string. The induced metric is \begin{equation} \label{metricainducida}
dh^2=\left(f^2(R)+\frac{\dot{R}^2}{f^2(R)}\right)dt^2 +\left(R^2+\frac{R'^2}{f^2(R)}\right)d\phi
^2 +2\frac{\dot{R}R'}{f^2(R)}dtd\phi.
\end{equation}

Integrating the first term in (\ref{accion}) in the radial variable we find, \begin{equation} \label{BB}  I = \int dtd\phi\left[\frac{\epsilon}{4} (l^2-R^2)-\mu \sqrt h\right].
\end{equation}
We  now  insert  the de Sitter metric (\ref{Ds}) with cosmological constant $1/l^{2}$ and the induced metric (\ref{metricainducida}) in (\ref{BB}). The resulting action is function of the position of the string only, $R(r,\phi)$. In order to compute the number of negative eigenvalues of the operator $\mathcal{O}$  we will expand the around the classical -static- solution of radius $R_0$ as \begin{equation}\label{expand} R = R_0 + \Delta(t,\phi) \end{equation} up to second order in $\Delta$ such that \begin{equation}\label{Be} I = I_0 + I_1 + I_2 + {\cal O}(\Delta^3).
\end{equation}
The first term, $I_0=-B$, was computed in the previous section and it is recovered here. The second term, \begin{equation}\label{ecmov} I_{1} = -\Delta\left(\frac{\epsilon}{2} f(R_{0})R_{0}l^2+\mu(l^2-2R_0^2)\right),
\end{equation}
vanishes when the equations of motion are used as expected. It is easy to see that $I_{1}$ is equivalent to  Eq.(\ref{radio}) in the limit $G\rightarrow 0$.

The operator $\mathcal{O}$ can be read from, \begin{equation} \label{Scuadratica} I_2=-\Delta ^2\left(\frac{\epsilon}{2}+\frac{\mu r}{f^3}\left(2\frac{R_0^2}{l^4}- \frac{3}{l^2}\right)\right)-\dot{\Delta}^2\left(\frac{\mu
R_0}{f^3}\right)-\Delta'^2\left(\frac{\mu}{R_0f}\right),
\end{equation}
which using Eq.(\ref{ecmov}) may be rewritten as \begin{equation}\label{casioper}
I_2=\frac{2\mu}{f^3 R_0}(\Delta^2-R_0^2\dot\Delta^2-f^2\Delta'^2).
\end{equation}
Making the change of variable $t=\frac{\beta}{2\pi}\tau$, with $0<\tau<2\pi$ and $\beta=2\pi l$ the inverse of temperature of the cosmological horizon, the operator (\ref{oper}) is given by \begin{equation}\label{operador2} {\mathcal{O}} = -\frac{4\mu}{f^3 R_0}\left[1+\left(\frac{2\pi R_0}{\beta}\right)^2\partial_\tau^2+ f^2\partial_\phi^2\right].
\end{equation}
The eigenvalues of ${\cal O}$ are proportional to the positive constant $4\mu/f^3 R_0$. As we are only interested in the sign of the eigenvalues, we will only consider the ones associated with the operator in square brackets of (\ref{operador2}).  They are \begin{equation}\label{autovalor}
\lambda_{nm}=-1+n^2 \frac{R_0^2}{l^2}+m^2f^2=-1 + m^2 + \frac{R_0^2}{l^2}\left(n^{2}-m^{2}\right).
\end{equation}
where $n,m\in \mathbb{Z}$.

Recalling that $R_{0}^{2}< l^{2}$, it is direct from Eq.(\ref{autovalor}) to notice that there are
five negative eigenvalues: $\lambda_{0 0}$,   $\lambda_{\pm 1 0}$ and $\lambda_{0\pm 1}$.
Furthermore, there are four null eigenvalues given by $\lambda_{\pm 1 \pm 1}=0$. The presence of these null eigenvectors is due to the $SO(4)$ symmetry of the background geometry. Indeed, from the six symmetry generators, only four rotate the string. The remanning two rotate it on itself, corresponding to reparameterizations of the worldsheet.  In appendix \ref{fromSymmetries}  we show that it is also possible to count the number of negative eigenvalues making use these symmetries.
It is worth noticing that this result is in complete agreement with the mathematical analysis carried out in \cite{urbano} where the number of negative eigenvalues of the operator associated with area of a two-torus immersed on $S^{3}$ is, at least,  five.

If we accept an arbitrary value of $\beta$  then we may reduce the number of negative eigenvalues to only three. In fact, it is easy to see that if \begin{equation}  \beta<2\pi R_0 \label{beta} \end{equation} the only eigenvalues which are always negative are $\lambda_{0 0}$  and $\lambda_{0\pm 1}$. This means that if the temperature is high enough, \textit{i.e.}, if one puts the system in contact with a heat reservoir with a temperature satisfying (\ref{beta}), then one may reduce, in the best case, the number of negative eigenvalues to three. This, however, amounts to admit a conical singularity at the cosmological horizon.

%%%%%%%%%%%%%%%%%%%%%%%%%%%%%%%%%%%%%%%%%%%%%%%%%%%%%%%%%%%%%%%%%%%%%
\section{Thermalon interacting with gravity. Analysis of the prefactor }\label{main} %%%%%%%%%%%%%%%%%%%%%%%%%%%%%%%%%%%%%%%%%%%%%%%%%%%%%%%%%%%%%%%%%%%%%
%%%%%%%%%%%%%%%%%%%%%%%%%%%%%%%%%%%%%%%%%%%%%%%%%%%%%%%%%%%%%%%%%%%%%

Now we let gravity  interact dynamically with the string.  We are interested in computing the one-loop correction to the semiclassical approximation of the path integral corresponding to the following Hamiltonian action, \begin{equation} \label{accionH} I=\int_{\mathcal{M}} (\dot g_{ij}\pi^{ij}+\dot A_{ij}P^{ij}-N^{\perp} \mathcal{H}_{\perp}-N^i\mathcal H_i -\lambda_iG^i)d^3x
+ \int_{\Sigma} d^2x  p_\mu \dot{y}^\mu + \pi {\cal A}.
\end{equation}
Here $g_{ij}$, $\pi^{ij}$ are the canonical variables describing the gravitational field, $A_{ij}$, $P^{ij}$ are the ones corresponding to the electromagnetic fields, while the pair $y_\mu$, $p^\mu$ correspond to the string degrees of freedom. ${\cal H}^i$, ${\cal H}^\perp$ are the gravitational constraints, and $G^i$ the electromagnetic Gauss constraints. $N^i, N^\perp, \lambda_i$ are Lagrange multipliers. The last term is ${\cal A}/4G \hbar$, ${\cal A}$ being the area of the cosmological horizon, and it is required for ensuring invariance under change of coordinates (We use the same conventions adopted in  \cite{Banados:1993qp}).

The action (\ref{accionH}) contains gauge degrees of freedom. One way to perform the path integral is to fix the gauge and work in the reduced phase space. This amounts to include a delta function on a set of gauge fixing conditions, $$ \chi^\mu = 0 $$ and to complement the integration measure with  the determinant of the operator $\{H_\mu, \chi^{\nu}\}$(see, for example, \cite{HTBook}).
Here we run into difficulties. In the gauge we have chosen (see appendix \ref{ap1}), the determinant in question turns out to be untractable. On the other hand, this gauge has the advantage of allowing us to completely solve the constraints and write the reduced action explicitly. We, therefore, will work with the reduced action and will assume that the new measure will not affect the number of negative eigenvalues we are counting.

The metric may be written in an ADM form as follows, \begin{equation} ds^2=N^2dt^2 + d\rho^2+\gamma^2(d\phi+N^\phi dt)^2.
\end{equation}
Here we have chosen the following gauge fixing conditions \begin{equation}
    g_{\rho\rho} = 1 \ \ \ \ \ \ \ \ \     g_{\rho\phi}=0 \label{grr}.
\end{equation}
To completely fix the diffeomorphism invariance we still have to choose a third condition. We require $\pi^{\phi\phi}= 0$. These gauge fixing conditions were first used to completely solve the constraints  in \cite{teitelfix}. We are going to gauge fix the electromagnetic field as well by requiring $A_{\phi\rho}=0$.

The constraints ${\cal H}^i={\cal H}^\perp=0$, whose explicit form is included in appendix \ref{ap1}, along with the gauge conditions form a set of second class constraints that may be completely solved outside the strings, in terms of four functions of $(t,\phi)$. At each side of the string, we get \begin{eqnarray} \gamma^2&=& r_-^2
\cos\left(\frac{\rho-\rho_0}{l}\right)^2+r_+^2 \sin\left(\frac{\rho-\rho_0}{l}\right)^2\nonumber\\
\pi^\rho_\phi &=& \frac{r_+r_-}{l\kappa}\\ \pi^\rho_\rho &=& -\frac{1}{\kappa}\frac{\partial}{\partial\phi}\arctan\left(
\frac{r_-}{r_+}\tan\left(\frac{\rho-\rho_o}{l}\right)\right) +\Pi.\nonumber \end{eqnarray} The four functions $r_+(t,\phi)$, $r_-(t,\phi)$, $\rho_0(t,\phi)$, $\Pi(t,\phi)$ are arbitrary.
However, in general, they will be different at each side of the string. The way they jump across it is determined by the constraints, as we shall see below.

The conservation of the gauge conditions (\ref{grr}) in time, $\dot{\pi}^{\phi\phi}=\dot{g}_{\phi\rho}=\dot{g}_{\rho\rho}=0$,  give rise to a set of consistency conditions, \begin{eqnarray} 0 &=& N^\rho_{,\rho} \ \ , \nonumber\\ 0 &=& N^\rho_{,\phi} +
\gamma^2 N^{\phi}_{,\rho} + 2\kappa\frac{\pi^{\rho}_{\phi}}{\gamma}N\
 \ , \label{consist}\\
0 &=& \frac{1}{\kappa\gamma}N_{,\rho\rho}- \frac{1}{\gamma}\left(\kappa
\frac{(\pi^{\rho}_{\phi})^2}{\gamma^4}- \frac{1}{l^2\kappa}\right) N- \frac{2\pi^{\rho}_{\phi}}{\gamma^2}N^\phi_{,\rho} +  \mu \int N^2 \frac{d}{d\gamma^2}(\sqrt h + h^{tt})\delta^{(3)}(x-y)d^2\sigma.\nonumber
\end{eqnarray}
These equations restrict the possible values of the Lagrange multipliers $N$, $N^i$. The most general solution  may be given in terms of four arbitrary functions of $(t,\phi)$: $u(\phi,t)$, $v(\phi,t)$, $w(\phi,t)$ and $n^\rho(\phi,t)$, \begin{eqnarray} N&=&\frac{1}{\gamma}\left [ u
\cos\left(\frac{\rho-\rho_0}{l}\right) \sin\left(\frac{\rho-\rho_0}{l}\right) + v \left( r_+^2 \sin\left(\frac{\rho-\rho_0}{l} \right)^2 -r_-^2\cos\left(\frac{\rho-\rho_0}{l}\right)^2\right)
\right] \\ N^\phi&=&w+\frac{1}{r_+\gamma^2}\cos\left(\frac{\rho-\rho_0}{l}\right)\left[ 2 r_+^2
r_- v \sin\left(\frac{\rho-\rho_0}{l}\right)+ur_-\cos\left(\frac{\rho-\rho_0}{l}\right) \right]\\ N^\rho &=& n^\rho.
\end{eqnarray}
where again the fields may jump across the string.

To compute the operator associated with the second variations of the action, we consider the most general variation of the metric and string position  respecting the constraints above.

The variational principle we use is adapted for fixing $\gamma$, $N^i$ and $N$ at the boundary $\rho=0$. We  fix $\gamma=0$, $N^\phi=0$, $N^\rho=0$ and $N=1$, therefore the metric in the $(+)$-region has \begin{eqnarray} {\gamma}_{(+)}&=& {r_+}^{(+)}\sin\left(\frac{\rho}{l_+}\right),\label{ga+}\\
N^\phi_{(+)}&=&0,\label{Np+}\\
N_{(+)}&=&\cos\left(\frac{\rho}{l_+}\right)+{v}^{(+)}{r_+}^{(+)}\sin\left(\frac{\rho}{l_+}\right).\label{N+}
\end{eqnarray}
The fields ${r_-}^{(+)} $ and ${v}^{(+)}$ remain arbitrary and we have to sum over them in the path integral.

The metric functions in the $(-)$-region have six functions determined by the three continuity conditions for $\gamma$, $N$ and $N^\phi$ and the three jumping condition (\ref{c1}), (\ref{c2}),
(\ref{c3})  for its radial derivatives across the string.  This is, in general, a very difficult task. However, we only need to consider the second order variations of the action, therefore we make an expansion of all functions around the thermalon solution. For the position of the string we write, \begin{eqnarray}
\bar\rho    &=&\bar\rho^{(o)}+\lambda\bar\rho^{(1)},            \label{serieR}\\
\dot{\bar\rho}&=&\lambda\dot{\bar\rho}^{(1)},         \label{serieRd}\\
{\bar\rho}'&=&\lambda{\bar\rho}'^{(1)}. \label{serieRp} \end{eqnarray} In terms of this parameter we write the variation of the remaining fields to order $\lambda^2$,

\begin{eqnarray}
{r_+}^{(+)}   &=& l_+  +  \lambda{r_+}^{(1+)} + \lambda^2{r_+}^{(2+)}+ {\cal O}(\lambda^3),          \label{serier-+}\\
{v}^{(+)}      &=&           \lambda {v}^{(1+)}  + \lambda^2{v}^{(2+)}+ {\cal O}(\lambda^3), \\
{r_-}^{(-)}    &=&           \lambda {r_-}^{(1-)} + \lambda^2{r_-}^{(2-)} + {\cal O}(\lambda^3)    ,                  \label{serier--}\\
{r_+}^{(-)}  &=& \sqrt{M} l_-  + \lambda{r_-}^{(1-)} + \lambda^2{r_-}^{(2-)}+ {\cal O}(\lambda^3),            \label{serier+-}\\
{u}^{(-)}      &=& C\sqrt{M}l_-+\lambda{u}^{(1-)}+\lambda^2{u}^{(2-)}+ {\cal O}(\lambda^3) ,                   \label{serieu-}\\
{v}^{(-)}      &=&           \lambda {v}^{(1-)}  + \lambda^2{v}^{(2-)}+ {\cal O}(\lambda^3) ,                           \label{seriev-}\\
{w}^{(-)}      &=&           \lambda {w}^{(1-)}  + \lambda^2{w}^{(2-)}+ {\cal O}(\lambda^3),                          \label{seriew-}\\
{\rho_o}^{(-)}   &=& \rho_o + \lambda{\rho_o}^{(1-)} + \lambda^2{\rho_o}^{(2-)}+ {\cal
O}(\lambda^3). \label{serierho0}
\end{eqnarray}

We now solve the equations order by order in $\lambda$. We end up with five arbitrary functions, $\bar\rho^{(1)}$, $\dot{\bar\rho}^{(1)}$, ${\bar\rho}'^{(1)}$, ${r_+}^{(1+)}$ and  ${v}^{(1+)}$ in terms of which all the other fields are determined. Each of these five functions must be integrated in the path integral.

Finally, the action evaluated on the \textit{Maxwell} equations (\ref{AccionOnMaxwell}) is given by

\begin{eqnarray}
I &=& -E_+^2\int^{\bar\rho}_0 N_{(+)}\gamma_{(+)}d\rho dtd\phi - E_{-}^{2}\int_{\bar\rho}^{\rho_{--}} N_{(-)}\gamma_{(-)}d\rho dtd\phi-\mu\int_{\Sigma} \sqrt{N^2{\bar\rho}'^{2} - N^2\gamma^2-\dot{\bar\rho}\gamma^2
}dtd\phi \nonumber\\
   &+& \frac{\pi}{\beta}\int\gamma_{(-)}dtd\phi+E_+^2\int^{\rho_{++}}_0
N_{(+)}\gamma_{(+)}d\rho dtd\phi+\frac{1}{2} \int (N^{(+)}\gamma_{(+),\rho}-1) dtd\phi,\label{AcOnMaxwell} \end{eqnarray} where the third term is evaluated on $\rho=\bar\rho$, the fourth term is evaluated on $\rho=\rho_{--}$ (the cosmological horizon), the last term is evaluated at $\rho=0$ and $\rho_{++}=\pi l_+/2$ is where $\gamma_{+}$ is maximum. The fourth term, on shell, is equal to ${\mathcal{A}}/4$ where $\mathcal{A}$ is the area of the cosmological horizon. Replacing the functions as series of $\lambda$ in (\ref{AcOnMaxwell}) we can express the action in the following form, \begin{equation} I=I_o+\lambda I_1+\lambda^2 I_2, \end{equation} where the term $I_2$ can written as, \begin{equation}\label{accionSegOr}
I_2=-\frac{1}{2}\left(\bar\rho^{(1)}(a+b\partial_t^2+c\partial_\phi^2)\bar\rho^{(1)}+U\,x\,U+V\,y\,V\right).
\end{equation}
Here the parameters $a, b, c, x$ and $y$ are functions of $l_+$, $l_-$ and $\mu$. The functions $U$ and $V$ are linear combination of $\bar\rho^{(1)}$, ${r_+}^{(1+)}$ and ${v}^{(1+)}$.

The operator $\mathcal{O}$ can be written in this case as \begin{equation} {\mathcal{O}} = \begin{pmatrix}
         a+b\partial_t^2+c\partial_\phi^2 & 0 & 0\\
          0 & \lambda_1 & 0  \\
         0 & 0 & \lambda_2 \\
              \end{pmatrix}.
\end{equation}

The eigenvalues of this operator are \begin{eqnarray} \lambda_1=x, \lambda_2=y\\ \lambda_{nm}=a-b\left(\frac{2\pi}{\beta}\right)n^2-cm^2. \nonumber \end{eqnarray} where $n,m\in \mathbb{Z}$.

We study numerically the quantities $a, b, c, x$ and $y$ for different values of the parameters.
The result of those numerical computations yields the minimum number of negative eigenvalues is twelve. This result proves that the presence of gravity increases the number of negative eigenvalues.

\section{Conclusions}

In this work we have analyzed the thermalon process in three spacetime dimensions, centered in the problem of the number of negative eigenvalues of the operator $\mathcal{O}$ that arises from second order perturbations around the classical solution.

When gravitational back reaction is neglected, the operator has five negative eigenvalues. We point out that if the fixed geometry represents a de Sitter geometry in thermal contact with a reservoir with high enough temperature, then the number of eigenvalues may be reduced to three.
This, however, amounts to accept a conical singularity in the background geometry.

Since the presence of more than one negative eigenvalue seems to rule out the interpretation of the thermalon as the indication of metastability also in this work we explored the influence of gravity expecting that this would reduce the number of negative eigenvalues. The result was the opposite and the presence of gravity seems to increase the number of negative eigenvalues.

In this work we neglected the influence of the measure in the path integral. Although the correct treatment of the measure may reduce the number of negative eigenvalues, as  it does for the \textit{conformal mode} of gravity in \cite{MM}, in our case it is unlikely that this could account to eliminate a plethora of negative eigenvalues. Therefore, we argue that the inclusion of the gravitation backreaction not only does not reduce the number of negative eigenvalues but most probably it actually increases them, spoiling the standard interpretation, \textit{a l\`{a}} Coleman, of the thermalon in the context of thermal decay rates.

\vspace{1in}

\section*{Acknowledgments}
 We would like to to thank G. Barnich, S. Carlip, J. Edelstein, D. Pons and R. Portugues for useful discussions.
This work was partially funded by grants FONDECYT 1040202, 1051084,1051064 and DI 06-04, 26-05.
(Universidad Andr\'es Bello). A. M. would like to thank CONICYT for its financial support. R.A.
would like to thank Abdus Salam International Centre for Theoretical Physics (ICTP)for its support.

\appendix

\section{Analysis from the symmetries}\label{fromSymmetries} %%%%%%%%%%%%%%%%%%%%%%%%%%%%%%%%%%%%%%%%%%%%%%%%%%%%%%%%%%%%%%%%%%%%%

In this appendix the result in Eq.(\ref{autovalor}) will be reobtained starting form the symmetries of the problem. This is in analogy of what was done by Garriga in \cite{garriga}. We will study the fluctuations of the string around the thermalon solution $r=R_0$. The operator that represents the second variations in the action must be covariant on the string worldsheet. This implies that it must be proportional to $-\nabla^2+C$, where $-\nabla^2$ is the laplacian on the worldsheet and $C$ is a constant. Now, \begin{equation}\label{laplacian} \nabla^2 = \frac{1}{f^2}\partial^2_t + \frac{1}{R_0^2}\partial^2_\phi.
\end{equation}
We define a dimensionless coordinate $\tau=\frac{2\pi t}{\beta}$, so that $\tau\in[0,2\pi]$, and the operator is proportional to \begin{equation}\label{operator} -\frac{(2 \pi)^2}{f^2\beta^2}\partial^2_\tau - \frac{1}{R_{0}^2}\partial^2_\phi + C.
\end{equation}
The eigenvalues in the space of $2\pi$-periodic functions in $\phi$ and $\tau$ are, therefore, proportional to \begin{equation}\label{eigenvalues} \lambda_{nm} \propto \hat{\lambda}_{nm} =
\frac{(2 \pi)^2n^2}{f^2\beta^2} +\frac{m^2}{R_{0}^2}+ C, \end{equation} where $n,m\in \mathbb{Z}$.
In order to determine $C$ we note first that this constant does not depend on $\beta$. This is because the global parameter $\beta$ appears only through the redefinition of the time coordinate, and therefore the Lagrangian density, which is a local quantity, cannot depend on it.  Hence, $C$ may be computed for some particular value of $\beta$ where the answer is known. In the case in which $\beta=2\pi l$, the geometry is a 3-sphere (globally Euclidean de Sitter) and one knows that
4 zero modes must show up. This is because the 3-sphere has six symmetries, two of which leave the position of the thermalon, $r=R_{0}$, invariant (representing reparameterizations of the worldsheet of the string). The other four should give rise to these zero eigenvalues. Note that the corresponding killing vectors must have periodicity $2\pi$, for they correspond to rotations, and therefore they must have $n^2=m^2=1$. This are precisely the four modes we need, and therefore \begin{equation}\label{zero} \hat{\lambda}_{\pm 1\pm 1} = 0 = \frac{(2 \pi)^2}{f^2\beta^2}
+\frac{1}{R_{0}^2}+ C, \end{equation} which implies
\begin{equation}
C = -\frac{1}{R_{0}^2 f^2(R_0)} . \label{dslimit} \end{equation} As a double check let us take another limit where we know the number of 0-modes: flat spacetime at finite temperature. This is 3 dimensional flat space with the Euclidean time coordinate identified
with periodicity $\beta$ which correspond to the inverse temperature.   The complete
six-dimensional group of rotations and translation is broken to translations plus the rotation in the plane orthogonal to $t$. The other two rotations do not respect the identification (do not commute with $\partial_t$). From the four remaining symmetries, two are, again, just reparameterizations of the string worldsheet, and therefore we are only left with  two spacial translations, which correspond to killing vectors independent of $t$ that have periodicity $2\pi$ in \textit{space}. We expect that in this case the eigenvalues with $n=0, m=\pm 1$ vanish. In fact, \begin{equation} \lim_{l\rightarrow\infty}\hat{\lambda}_{0\pm 1} = \frac{1}{R_{0}^2}+\lim_{l\rightarrow\infty}C = 0.
\end{equation}
Now plugging (\ref{dslimit}) back in (\ref{operator}) we obtain precisely our result
(\ref{operador2}) of Section (\ref{fixedback}).

%%%%%%%%%%%%%%%%%%%%%%%%%%%%%%%%%%%%%%%%%%%%%%%%%%%%%%%%%%%%%%%%%%%%%%%%
%%%%%%%%%%%%%%%%%%%%%%%%%%%%%%%%%%%%%%%%%%%%%%%%%%%%%%%%%%%%%%%%%%%%%%%%%
\section{The Hamiltonian formulation of the Thermalon} \label{ap1} %%%%%%%%%%%%%%%%%%%%%%%%%%%%%%%%%%%%%%%%%%%%%%%%%%%%%%%%%%%%%%%%%%%%%%%%%%%
%%%%%%%%%%%%%%%%%%%%%%%%%%%%%%%%%%%%%%%%%%%%%%%%%%%%%%%%%%%%%%%%%%%%%%%%%%%

The Hamiltonian action is given by (\ref{accionH}).
 On the constraint surface, the canonical action is simply \begin{equation}  \label{AccionHamEval}  I=\int_{\Sigma} p_\mu\dot{y^\mu} + \pi {\cal A}.
\end{equation}
 The canonical momentum $p_\mu$ conjugate to the string is is given by \begin{equation} \label{momentum} p_\mu=-\mu\sqrt h g_{\mu\nu}(h^{00}\dot y^\nu+h^{0i}y'^\nu)-q A_{\mu\nu}y'^\nu.
\end{equation}
Replacing Eq.(\ref{momentum}) in Eq.(\ref{AccionHamEval}) this action may be written as, \begin{equation} \label{Ac} I=-\mu\int_{\Sigma} d^2\sigma \sqrt h - q\int_{\Sigma} A  + \pi {\cal A}.
\end{equation}

The action principle  (\ref{accionH}) must be supplemented with information on the fields which are going to be fixed on the boundary. In the present case the boundary is defined to be at $r=0$, mimicking what was done in \cite{GHTW}. When varying the action one gets the following boundary terms at $\rho=0$ $$ -\frac{1}{2}\int_{\partial \mathcal{M}}\left( N\delta\gamma_{,\rho} - N_{,\rho}\delta\gamma \right) + \int_{\partial \mathcal{M}} N^i\delta \pi_i^\rho .
$$
We see that the action is ready to fix $\gamma$ and $\gamma_{,\rho}$ and $\pi^\rho_i$ on the boundary. We are interested in fixing $N=1$ instead of $\gamma_{,\rho}$, so that the inverse temperature $\beta=(t_2-t_1)$ will be fixed, as it should be in the canonical ensemble. To accomplish that we may add to the action the boundary term $$ \frac{1}{2}\int_{\partial \mathcal{M}}(N\gamma_{,\rho} - K), $$ where $K$ is an arbitrary constant. In order to compute $K$ we consider the situation when there is no string and we have a particle singularity at $\rho=r=0$. In that case the boundary term on shell is given $\beta(1-M-K)$, which shows that $K=1$. In fact, the complete action on-shell is $$ \beta(-M) - \pi{\cal A} .
$$
Recalling that the action in the canonical ensemble represents the free energy, and $\pi{\cal A}$ the entropy, we find the internal energy is $-M$. This is in agreement with what was done in \cite{Gomberoff:2003ea}. The mass will be $+M$ when the boundary is set at the cosmological horizon.

After taking the boundary terms into account, the action takes the form $$ I = -\mu\int_{\Sigma} d^2\sigma \sqrt h - q\int_{\Sigma} A +  \pi {\cal A}+ \frac{1}{2}\int_{\partial \mathcal{M}}(N\gamma_{,\rho} - 1).
$$

The electromagnetic constraint in the action requires $P^{ij}$ to be fixed on the boundary. In our case, we want $P^{ij}=E_+ \epsilon^{ij}$ at the boundary, where $\epsilon^{ij}$ is the Levi-Civita tensor on the constant $t$ slice. The constraints require that $E_+$ is independent of $\phi$.
Furthermore, we want to describe a time-independent cosmological constant and therefore we choose $E_+$ to be a constant. Gauss' constraint $G^{i}$ will completely determine the value of the field $E$ everywhere. We hence find that on the constraint surface we are allowed to use the Maxwell equations $d\,^{\ast} F=-{}^*\! J$. We may then write, on the constraint surface,

\begin{equation} q\int_{\Sigma} A = \int_{\mathcal{M}} {}^*\!J \wedge A  = -\int_{\mathcal{M}} d{}^*\!F \wedge A  = \int_{\mathcal{M}} {}^*\!
F\wedge F-\int_{\partial \mathcal{M}}{}^*\! F\wedge A, \end{equation} where $\partial {\mathcal{M}}$ is the boundary of $\mathcal{M}$ at $r=0$. The action is \begin{equation} I=-\mu \int_{\Sigma} \sqrt h d^2\sigma- \int_{\mathcal{M}} {}^*\! F\wedge F +\int_{\partial \mathcal{M}} {}^*\! F\wedge A + \pi {\cal A}+\frac{1}{2}\int_{\partial \mathcal{M}}(N\gamma_{,\rho} - 1), \end{equation} which may also be written as, \begin{equation}\ \label{AccionOnMaxwell} I=-\mu \int_{\Sigma} \sqrt h d^2\sigma-\int_{\mathcal{M}} \sqrt g E^2 d^3x +\int_{\partial \mathcal{M}}{}^*\! F\wedge A +\pi {\cal A}+\frac{1}{2}\int_{\partial \mathcal{M}}(N\gamma_{,\rho}
- 1).
\end{equation}

We finish this section by computing the form of the gravitational degrees of freedom left after solving the constraints. The Hamiltonian constraints contain a pure gravitational piece and
a matter term,   ${\mathcal H}={\mathcal H}^{Grav}+{\mathcal H}^{Mat}$. In the present case
\begin{eqnarray}
\label{constrGrav} {\mathcal H}^{Grav}_{\perp} &=& 2\kappa\gamma^{-3}(\pi^\rho_\phi)^2 - \frac{2}{l^2\kappa}\gamma -\frac{2}{\kappa}\gamma_{,\rho\rho}\ \  , \\ {\mathcal H}^{Grav}_{\rho} &=& -2\left(\pi^\rho_{\rho,\rho} + \{\gamma^{-2}\pi^\rho_\phi\}_{,\phi}\right) \ \ , \\ {\mathcal H}^{Grav}_{\phi} &=& -2 \pi^\rho_{\phi,\rho}.
\end{eqnarray}
The matter part is given by
\begin{equation}\label{constrMat1}
{\mathcal H}_\perp^{Mat} =\sqrt g T_{\perp\perp} \textrm{ and } {\mathcal H}_i^{Mat} = \sqrt g T_{\perp i}, \end{equation} where $\sqrt g=\gamma$,  $T_{\perp\perp}=T^{\mu\nu}n_\mu n_\nu$, $T_{\perp i}={T^{ \mu }}_i n_\mu$ and $n_\mu=(N,0,0,0)$ is the unitary timelike vector normal to the $t=$constant surfaces. The string may be parameterized  by $\rho=R(t,\phi)$. The induced geometry on it is  \begin{equation}  d\sigma^2=(N^2+\dot R^2+\gamma^2N^{\phi2})dt^2+2(\dot R R'+\gamma^2N^\phi)dtd\phi+(R'^2+\gamma^2)d\phi ^2.
\end{equation}
The energy momentum tensor produced by the string and the electromagnetic field is given by \begin{equation} \label{Tmunu} T^{\mu\nu}= \frac{1}{2} E^2 g^{\mu\nu}+\mu\int \frac{\sqrt h }{\sqrt{g^{(3)}} }\frac{\partial y^\mu}{\partial \sigma^a}\frac{\partial y^\nu}{\partial \sigma^b} h^{a b}\delta^{(3)}(x-y)d^2\sigma.
\end{equation}
Now, using (\ref{constrMat1})-(\ref{Tmunu}), we may compute the matter part of the constraints, \begin{eqnarray} {\cal H}^{Mat}_{\perp} &=& \gamma N^2\left[\frac{E^2}{2N^2}+\mu\int \frac{\sqrt h}{N\gamma}h^{tt}\delta^{(3)}(x-y)d^2\sigma\right]\nonumber\\
{\cal H}^{Mat}_{\rho} &=& \gamma N\mu \int \frac{\sqrt h }{N\gamma}(h^{tt}\dot
R+h^{t\phi}R')\delta^{(3)}(x-y)d^2\sigma\\
{\cal H}^{Grav}_{\phi} &=& \gamma^3N\left[\frac{E^2g^{0\phi}}{2}+\mu\int\frac{\sqrt
h}{N\gamma}h^{t\phi}\delta^{(3)}(x-y)d^2\sigma\right].\nonumber
\end{eqnarray}

To find the jumping conditions we  integrate the constraints and consistency conditions across the string. We obtain \begin{eqnarray} \frac{2}{\kappa}\Delta \gamma,_\rho &=& \mu N \frac{R'^2+\gamma^2}{\sqrt h},\label{c1}\\ 2\Delta {\pi^\rho}_\rho &=& \frac{\mu \gamma^2}{\sqrt
h}(\dot{R}-R'N^\phi) \label{cons2}\\ 2\Delta{\pi^\rho}_\phi &=& -\gamma^2\mu\frac{(\dot RR'+N^\phi\gamma^2)}{\sqrt h}\label{cons3}.
\end{eqnarray}
For the  $N$'s, we note that the momentum ${\pi^\rho}_\phi$ in terms of the metric is given by \begin{equation}\label{pirhophi} {\pi^\rho}_\phi=\frac{\gamma^3}{2N}N^\phi,_\rho.
\end{equation}
Replacing (\ref{pirhophi}) in (\ref{cons3}) we obtain \begin{equation} \label{c2} \Delta N^\phi,_\rho=-\frac{\mu N}{\gamma}\frac{(\dot R R'+N^\phi \gamma^2)}{\sqrt h}.
\end{equation}

In the same way, (\ref{consist}) may  be integrated around the position of the string. This gives the discontinuity of the derivative of $N$, \begin{equation}\label{c3} \frac{1}{\kappa}\Delta N,_\rho=-N^2\gamma \mu \frac{d}{d\gamma^2}(\sqrt h h^{tt}) \end{equation} Note that in the expressions (\ref{c1}), (\ref{c2}) and (\ref{c3}) the right hand side is evaluated at the position of the string. This is not a problem, for  $N$, $\gamma$ and $N^\phi$ are continuous functions.

\bibliographystyle{jhep}

\end{document}